\documentclass[12pt]{article}
\usepackage{graphicx}

\newcommand{\be}{\begin{equation}}
\newcommand{\ee}{\end{equation}}
\newcommand{\ba}{\begin{eqnarray}}
\newcommand{\ea}{\end{eqnarray}}

\begin{document}

\begin{flushright}
{\small
SLAC--PUB--10216\\
JLAB-THY-03-191\\
WM-03-106\\
UMN-D-03-5\\
October 2003\\}
\end{flushright}

\begin{center}
{{\bf\LARGE Single-Spin Polarization Effects
and the \\ Determination of Timelike Proton \\[1.2ex] Form
Factors}\footnote{This work was supported in part by the Department of Energy
contracts DE-AC03-76SF00515 (S.J.B.), DE-AC05-84ER40150 (S.J.B. and C.E.C),
and DE-FG02-98ER41087 (J.R.H.);
by the National Science Foundation Grant PHY-0245056 (C.E.C);
and by the LG Yonam Foundation (D.S.H.).}}

\bigskip
{\it Stanley J. Brodsky \\
Stanford Linear Accelerator Center,
2575 Sand Hill Road, Menlo Park, CA 94025 \\
E-mail:  sjbth@slac.stanford.ed\\
and\\
Thomas Jefferson National Accelerator Facility\\
12000 Jefferson Avenue,
Newport News, VA 23606}

\bigskip

{\it Carl E. Carlson\\
Thomas Jefferson National Accelerator Facility\\
12000 Jefferson Avenue,
Newport News, VA 23606\\
and \\
Nuclear and Particle Theory Group, Physics Department\\
College of William and Mary,
 Williamsburg, VA 23187-8795\\
E-mail: carlson@physics.wm.edu}

\bigskip

{\it John R. Hiller\\
Department of Physics,
University of Minnesota-Duluth,
Duluth, MN 55812\\
E-mail:  jhiller@d.umn.edu}

\bigskip

{\it Dae Sung Hwang\\
Department of Physics,
Sejong University,
Seoul 143-747, Korea\\
E-mail: dshwang@sejong.ac.kr}

\end{center}

\vfill\newpage

\begin{center}
{\bf\large Abstract }
\end{center}

We show that measurements of the proton's polarization in $e^+  e^- \to
p \bar p$ strongly discriminate between analytic forms of models which
fit the proton form factors in the spacelike region.   In particular,
the single-spin asymmetry normal to the scattering plane measures the
relative phase difference between the timelike $G_E$ and $G_M$ form
factors. The expected proton polarization in the timelike region is
large, of order of several tens of percent.

\bigskip \bigskip




\section{Introduction}


The form factors of hadrons as measured in both the spacelike and
timelike domains provide fundamental information on the structure and
internal dynamics of hadrons. Recent measurements~\cite{perdrisat} of
the electron-to-proton polarization transfer in
$\overrightarrow e^- \, p \to e^- \, \overrightarrow p$
scattering at Jefferson Laboratory show that
the ratio of Sachs form factors~\cite{walecka} $G^p_E(q^2)/
G^p_M(q^2)$ is monotonically decreasing with increasing $Q^2=-q^2,$ in
strong contradiction with the $G_E/G_M$ scaling determined by the
traditional Rosenbluth separation method. The Rosenbluth method may in
fact not be reliable, perhaps because of its sensitivity to uncertain
radiative corrections, including two-photon exchange
amplitudes~\cite{guichon}. The polarization transfer
method~\cite{perdrisat,acg81} is relatively insensitive to such
corrections.

The same data which indicate
that $G_E$ for protons falls faster than $G_M$ at large spacelike $Q^2$
require in turn that $F_2/F_1$ falls more slowly than $1/Q^2$. The
conventional expectation from dimensional counting
rules~\cite{Brodsky:1974vy} and perturbative QCD~\cite{Lepage:1979za}
is that the Dirac form factor $F_1$ should fall with a nominal power
$1/Q^4$, and the ratio of the Pauli and Dirac form factors, $F_2/F_1$,
should fall like $1/Q^2$, at high momentum transfers. The Dirac form
factor agrees with this expectation in the range $Q^2$ from a few
GeV$^2$ to the data limit of 31 GeV$^2$. However, the Pauli/Dirac ratio
is not observed to fall with the nominal expected power, and the
experimenters themselves have noted that the data is well fit by
$F_2/F_1 \propto 1/Q$ in the momentum transfer range 2 to 5.6 GeV$^2$.

%
%
The new Jefferson Laboratory results make it critical to carefully
identify and separate the timelike $G_E$ and $G_M$ form factors by
measuring the center-of-mass angular distribution and by measuring
the polarization of the proton in $e^+ e^- \to p
\bar p$ or $p \bar p \to \ell^+ \ell^-$ reactions. The advent of high
luminosity $e^+ e^-$ colliders at Beijing, Cornell, and Frascati
provide the opportunity to make such measurements, both directly
and via radiative return.

Although the spacelike form factors of a stable hadron are real, the
timelike form factors have a phase structure reflecting the
final-state interactions of the outgoing hadrons. In general, form
factors are analytic functions $F_i(q^2)$ with a discontinuity for
timelike momentum above the physical threshold $q^2> 4 M^2.$ The
analytic structure and phases of the form factors in the timelike
regime are thus connected by dispersion relations to the spacelike
regime~\cite{baldini,Geshkenbein74,seealso}. The analytic form and
phases of the timelike amplitudes also reflects resonances in the
unphysical region $0 < q^2 < 4M^2$ below the physical
threshold~\cite{baldini} in the $J^{PC} = 1^{--}$ channel, including
gluonium states and di-baryon structures.

At very large center-of-mass energies, perturbative QCD factorization
predicts diminished final interactions in $e^+ e^- \to H \bar H$, since
the hadrons are initially produced with small color dipole moments.
This principle of QCD color transparency~\cite{Brodsky:xz} is also an
essential feature~\cite{Bjorken:kk} of hard exclusive $B$
decays~\cite{Beneke:2001ev,Keum:2000wi}, and thus needs to be tested
experimentally.

There have been a number of explanations and theoretically motivated
fits of the $F_2/F_1$ data.  Belitsky, Ji, and Yuan~\cite{belitsky02}
have shown that factors of $\log(Q^2)$ arise from a careful QCD
analysis of the form factors. The perturbative QCD form $Q^2 F_2/ F_1
\sim \log^2 Q^2$, which has logarithmic factors multiplying the nominal
power-law behavior, fits the large-$Q^2$ spacelike data well.
Others~\cite{ralston,miller} claim to find mechanisms that modify the
traditionally expected power-law behavior with fractional powers of
$Q^2$, and they also give fits which are in accord with the data.
Asymptotic behaviors of the ratio $F_2/F_1$ for general light-front
wave functions are investigated in~\cite{BHHK}. Each of the model forms
predicts a specific fall-off and phase structure of the form factors
from $ s \leftrightarrow t$ crossing to the timelike domain. A fit
with the dipole polynomial or nominal dimensional counting rule
behavior would predict no phases in the timelike regime.

As noted by Dubnickova, Dubnicka, and Rekalo, and by Rock~\cite{d}, the
existence of the $T-$odd single-spin asymmetry normal to the scattering
plane in baryon pair production $e^- e^+ \rightarrow B \bar B$ requires
a nonzero phase difference between the $G_E$ and $G_M$ form factors.
The phase of the ratio of form factors $G_E/G_M$ of spin-1/2 baryons in
the timelike region can thus be determined from measurements of the
polarization of one of the produced baryons.  We shall show that
measurements of the proton polarization in $e^+  e^- \to p \bar p$
strongly discriminate between the analytic forms of  models which have
been suggested to fit the proton $G_E/G_M$ data in the spacelike
region.

\section{Timelike Measures}

The center-of-mass angular distribution provides the analog of the
Rosenbluth method for measuring the magnitudes of various helicity
amplitudes.
%
%
The differential cross section for $e^- e^+ \rightarrow B \bar B$ when
$B$ is a spin-1/2 baryon is given in the center-of-mass frame by
\be                               \label{xsctn}
{d\sigma \over d\Omega} =
{\alpha^2 \beta  \over 4 q^2}
                            \ D \ ,
\ee
where $\beta = \sqrt{1-4m_B^2/q^2}$ and $D$ is given by
\be D =  |G_M|^2 \left(1+ \cos^2\theta \right) + {1\over \tau }\,
       |G_E|^2 \sin^2\theta  \ ;
\ee
we have used the Sachs form factors~\cite{walecka}
\ba G_M &=& F_1 + F_2 \ , \nonumber \\ G_E &=& F_1 + \tau F_2  \ ,
\ea
with $\tau \equiv {q^2 / 4 m_B^2} > 1$.

As we shall show, polarization observables can be used to
completely pin down the relative phases of the timelike form
factors. The complex phases of the form factors in the timelike
region make it possible for a single outgoing baryon to be
polarized in $e^- e^+ \rightarrow B \bar B,$ even without
polarization in the initial state.

There are three polarization observables, corresponding to
polarizations in three directions which are perhaps best called
longitudinal, sideways, and normal but often denoted $z$, $x$, and $y$,
respectively. Longitudinal ($z$) when discussing the final state means
parallel to the direction of the outgoing baryon. Sideways ($x$) means
perpendicular to the direction of the outgoing baryon but in the
scattering plane. Normal ($y$) means normal to the scattering plane, in
the direction of $\vec k \times \vec p$ where $\vec k$ is the electron
momentum and $\vec p$ is the baryon momentum, with $x$, $y$, and $z$
forming a right-handed coordinate system.

The polarization ${\cal P}_y$ does not require polarization in the
initial state and is~\cite{d}
\be                    \label{py} {\cal P}_y = {  \sin 2\theta
\, {\rm Im} G_E^*  G_M
      \over D \sqrt{\tau} }
                    =
      { (\tau - 1 )\sin 2\theta \, {\rm Im} F_2^* F_1
      \over D  \sqrt{\tau} }
                    \ .
\ee
%
%
The other two polarizations require initial state polarization.  If the
electron has polarization $P_e$ then~\cite{d}
\be                   \label{px} {\cal P}_x =
    -P_e {2 \sin\theta \, {\rm Re} G_E^* G_M \over D \sqrt{\tau} }
     \ ,
\ee
and
\be                   \label{pz} {\cal P}_z =  P_e {2  \cos \theta
|G_M|^2 \over D}  \ .
\ee
The sign of ${\cal P}_z$ can be determined from physical
principles. Angular momentum conservation and helicity conservation for
the electron and positron determine that ${\cal P}_z/P_e$ in the
forward direction must be $+1$, verifying the sign of the above formula.

The polarization measurement in $e^+ e^- \to  p \bar p$ will require a
polarimeter for the outgoing protons, perhaps based on a shell of a
material such as carbon which has a good analyzing power. However,
timelike baryon-antibaryon production can occur for any pair that is
energetically allowed. Baryons such as the $\Sigma$ and $\Lambda$ which
decay weakly are easier to study, since their polarization is
self-analyzing.

%
%
Polarization ${\cal P}_y$ is a manifestation of the T-odd observable
$\vec k \times \vec p \cdot \vec S_p$, with $\vec S_p$ the proton
polarization.  This observable is zero in the spacelike case, but need
not be zero in the timelike case because final state interactions can
give the form factors a relative phase.

Notice the factor $\sin 2\theta$.  Without polarization in the initial
state, and with single photon exchange, the only information
transferred to the final state is the total energy plus information for
the photon's polarization about the line---undirected---of the electron
and positron momentum in the initial state. Similarly, without using
polarization in the final state, we can use the undirected line of the
baryon momenta.  We can define a directed normal by taking a cross
product, providing a direction by rotating from the lepton to the
hadron direction through the smaller angle.  The observable is the dot
product of this directed normal with the baryon polarization.  At
$0^\circ$ or $90^\circ$, one cannot define a directed normal, hence one
cannot obtain nonzero polarization at these two angles, as reflected in
the $\sin 2\theta$ factor.

Any model which fits the spacelike form factor data with an analytic
function can be continued to the timelike region.  Spacelike form
factors are usually written in terms of $Q^2 = - q^2$.  The correct
relation for analytic  continuation can be obtained by examining
denominators in loop calculations in perturbation theory.  The
connection is $Q^2 \rightarrow q^2 e^{-i\pi}$, or
\be
\ln Q^2 = \ln (-q^2) \rightarrow \ln q^2 - i\pi  \ .
\ee
If the spacelike $F_2/F_1$ is fit by a rational function of
$Q^2$, then the form factors will be  relatively real in the timelike
region also.  However,  one in general gets a complex result from the
continuation.

More sophisticated dispersion
relation based continuations could give more reliable results, if there
is data also in the timelike region to pin down the magnitudes there.
So far, this is possible for the magnetic form factor
alone~\cite{baldini} but not for both form factors.


\section{Polarization in the timelike region}


We begin by selecting some existing fits to the spacelike data.
Since we are concentrating on the polarizations, which depend only
on the ratios of the form factors, we concentrate in turn on fits
to the ratio $F_2/F_1$, rather than fits to the individual form
factors. We attempt to present a representative selection of fits,
and refer to others that are similar to the ones included.

{\it Odd-$Q$ fits.} The JLab experimenters themselves note that the
polarization transfer data is well fit for $Q^2$ in the 2 to 5.6 GeV$^2$
region by
\be   \label{oddq}
{F_2 \over F_1} = {1.25 {\rm\ GeV} \over Q} \ .
\ee
There is theoretical work which obtains similar
forms~\cite{ralston,miller,moreoddq}.  Because of its simple
analyticity, this form becomes purely imaginary in the timelike
region.  Simply to get the right ratio at $Q^2 = 0$, we choose to
modify this form to
\be
\label{mododdq} {F_2 \over F_1} = \left( {1\over \kappa_p^2}
    + {Q^2 \over (1.25 {\rm\ GeV})^2} \right)^{-1/2} \ ,
\ee
where $\kappa_p = 1.79$ gives the anomalous magnetic moment of
the proton.  The numerical effect of the $1/\kappa_p^2$ term is hardly
noticeable for $Q^2$ above 2 GeV$^2$.

{\it Fits involving logarithms.}  A number of
authors~\cite{belitsky02,BHHK} have given fits to the $F_2/F_1$ data
which have the power law fall-off expected from QCD, with logarithmic
corrections that enable a good fit to the data.  Belitsky, Ji, and
Yuan~\cite{belitsky02} have motivated a form that has two powers of $\ln
Q^2$, and one of their fits is, with $\Lambda = 300$ MeV,
\be                                         \label{yuan} {F_2 \over F_1}
= 0.17 {\rm\ GeV}^2 \,
            {\ln^2 (Q^2 / \Lambda^2) \over Q^2} \ .
\ee
We give here an improved fit which matches the above
asymptotically and also matches low $Q^2$ data,
\be                                         \label{hiller}
{ F_2\over F_1}= \kappa_p {
    [1+(Q^2/0.791\,{\rm GeV}^2)^2 \ln^{7.1}(1+Q^2/4m_\pi^2)]
     \over [1+(Q^2/0.380\,{\rm GeV}^2)^3
     \ln^{5.1}(1+Q^2/4m_\pi^2)]
     }  \  .
\ee

This last form also contains the cut at the two-pion threshold in the timelike
region.

{\it Two-component fits.} In 1973, Iachello, Jackson, and
Lande~\cite{ijl} presented  a model for the nucleon form factors based
on a two-component, core and meson cloud, structure for the nucleon
with parameters fit to the then existing data.  The fit was updated by
Gari and Krumpelmann~\cite{gk} and later by Lomon~\cite{lomon}.
Iachello~\cite{iachello} has recently noted that one of the original
fits accords well with the newest JLab data.  Continued for timelike $q^2$, the fit is
\ba
\label{ijl} F_1 &=& {1\over 2}g \Bigg[ (1-\beta_\omega-\beta_\phi)
    - \beta_\omega {m_\omega^2 \over q^2 - m_\omega^2}
    - \beta_\phi {m_\phi^2 \over q^2 - m_\phi^2}
                        \nonumber \\
    &+& (1-\beta_\rho) - \beta_\rho
     {m_\rho^2 + 8\Gamma_\rho m_\pi/\pi \over
       q^2 - m_\rho^2 + (q^2 - 4m_\pi^2) \Gamma_\rho \alpha(q^2)/m_\pi}
    \Bigg] \ ,
                        \nonumber \\[2.5ex] F_2 &=& {1\over 2}g \Bigg[
(0.120+\alpha_\phi)
    {m_\omega^2 \over q^2 - m_\omega^2}
    - \alpha_\phi {m_\phi^2 \over q^2 - m_\phi^2}
                        \nonumber \\
    &-& 3.706 {m_\rho^2 + 8\Gamma_\rho m_\pi/\pi \over
       q^2 - m_\rho^2 + (q^2 - 4m_\pi^2) \Gamma_\rho \alpha(q^2)/m_\pi}
    \Bigg] \ ,
\ea
where
\ba
\alpha(q^2) &=& \left(q^2 - 4m_\pi^2 \over q^2 \right)^{1/2}
                        \nonumber \\
    &\times& \left\{{2\over\pi} \ln
            \left( \sqrt{q^2 - 4m_\pi^2}+\sqrt{q^2}
            \over 2 m_\pi \right) - i
    \right\} \ .
\ea
The function $g = g(q^2)$ cancels in expressions for
polarizations.  The parameters are $\beta_\rho = 0.672$,
$\beta_\omega=1.102$, $\beta_\phi=0.112$, $\alpha_\phi=-0.052$,
$m_\rho=0.765$ GeV, $m_\omega=0.784$ GeV, $m_\phi=1.019$ GeV, and
$\Gamma_\rho=0.112$ GeV.


\begin{figure}[htb]

\begin{center}
\includegraphics[height=4in,width=0.8\textwidth]{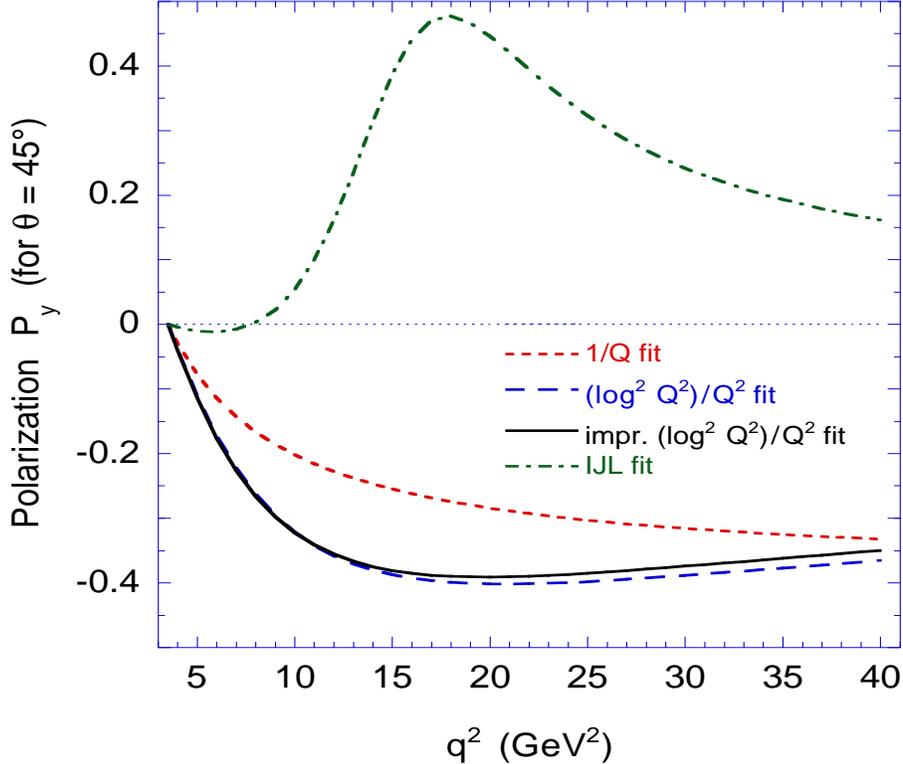}
\end{center}
\caption[*]{Predicted polarization ${\cal P}_y$ in the timelike region
for selected form factor fits described in the text.  The plot is for
$\theta = 45^\circ$.  The four curves are for an $F_2/F_1 \propto 1/Q$
fit, using Eq.~(\ref{mododdq}); the $(\log^2 Q^2)/Q^2$ fit of Belitsky
{\it et al.}, Eq.~(\ref{yuan}); an improved $(\log^2 Q^2)/Q^2$ fit,
Eq.~(\ref{hiller}); and a fit from Iachello {\it et al.},
Eq.~(\ref{ijl}).}
\label{figpy}
\end{figure}


Iachello~\cite{iachello} has also discussed extending the fits to the
timelike region, and finds a complex phase from two sources. One source
is a modification of the overall factor $g(q^2)$. The overall factor
has no effect on $G_E/G_M$ and no effect on quantities like
polarizations that only depend on ratios. The other source of phase is
the treatment of the rho widths. The phi and omega were approximated as
zero width, but IJL~\cite{ijl} found that rho-width contributions were
important for fitting data and incorporated a two-pion cut into an
effective width term in the rho propagator.  The extension to the
timelike region seen above is a straightforward and expected analytic
continuation of this term.

The expression for polarization ${\cal P}_y$, Eq.~(\ref{py}), leads to
results shown in Fig.~\ref{figpy}.  The polarizations are shown for
four fits listed above, and the polarizations are not small.  They are
very distinct from a purely polynomial fit to the spacelike data,
which gives zero ${\cal P}_y$.

\begin{figure}[htb]
\begin{center}
\includegraphics[height=4in,width=0.8\textwidth]{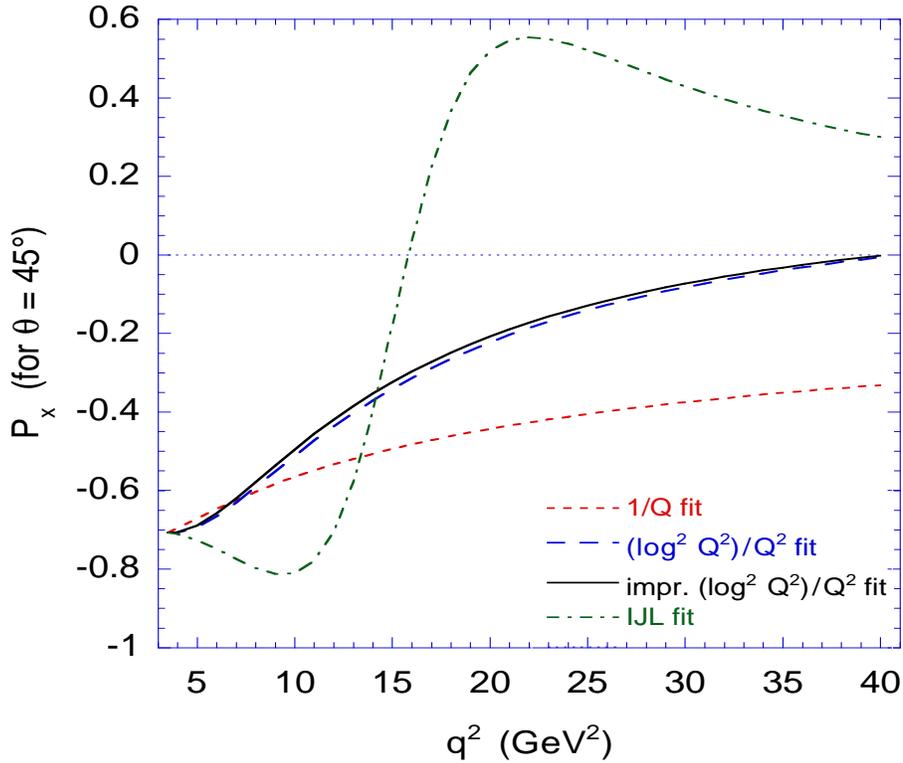}
\end{center}
\caption{The predicted polarization ${\cal P}_x$ in the timelike
region for $\theta=45^\circ$ and $P_e=1$.  The four curves correspond
to those in Fig.~\ref{figpy}.}
\label{fig:px}
\end{figure}

\begin{figure}[htb]
\begin{center}
\includegraphics[height=4in,width=0.8\textwidth]{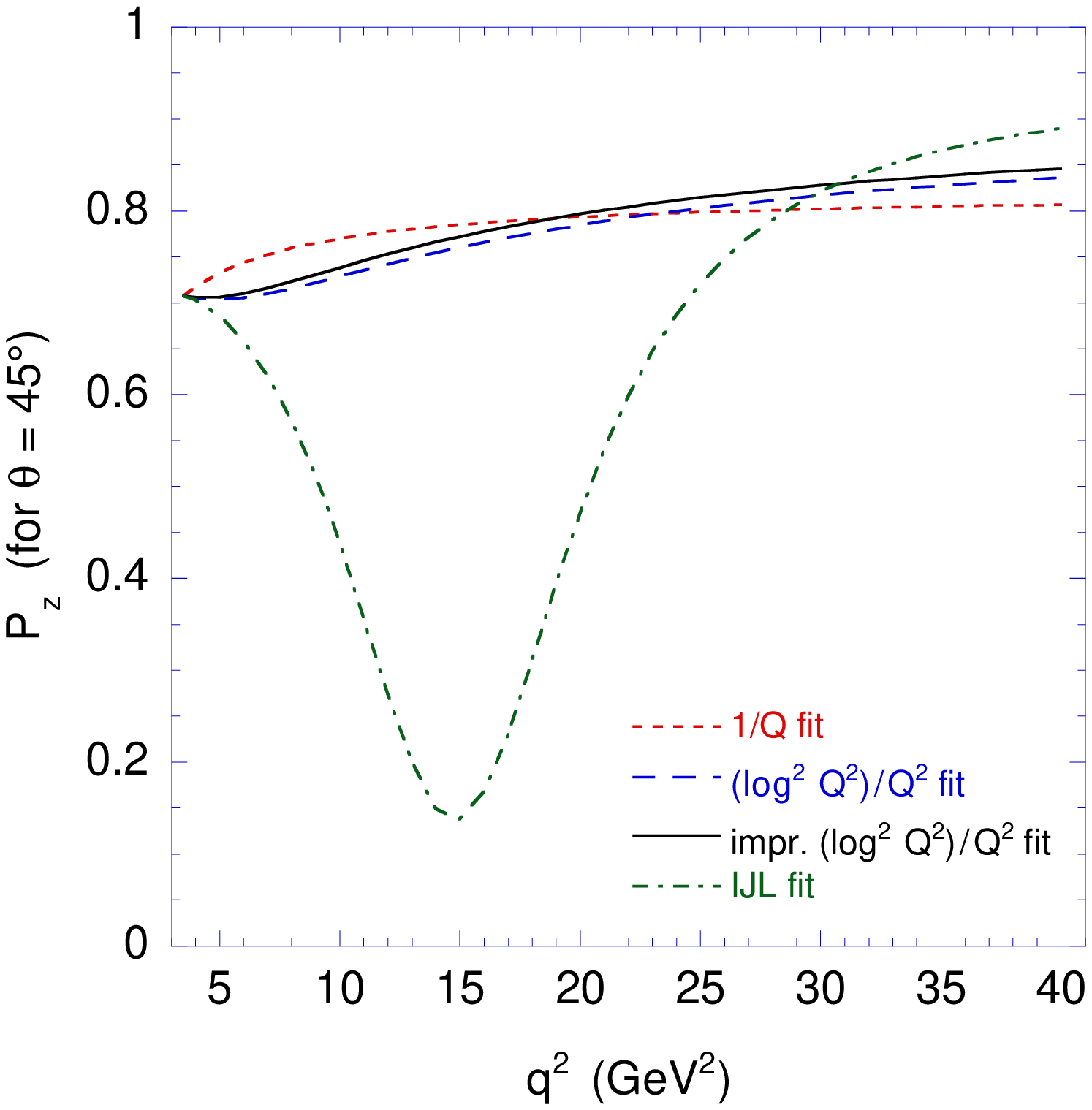}
\end{center}
\caption{The predicted polarization ${\cal P}_z$ in the timelike
region for $\theta=45^\circ$ and $P_e=1$.  The four curves
correspond to those in Fig.~\ref{figpy}. }
\label{fig:pz}
\end{figure}

The predictions for ${\cal P}_x$ and ${\cal P}_z$ are shown in
Figs.~\ref{fig:px} and~\ref{fig:pz}. Both figures are for scattering
angle $45^\circ$ and $P_e = 1$.   The phase difference
$(\delta_E-\delta_M)$ between $G_E$ and $G_M$ is directly given
by the ${\cal P}_y / {\cal P}_x$ ratio,
\be {{\cal P}_y \over {\cal P}_x} = {\cos\theta\over P_e}
    {{\rm Im\ } G_M^* G_E \over {\rm Re\ } G_M^* G_E}
    = {\cos\theta\over P_e} \tan(\delta_E-\delta_M)  \ .
\ee


\begin{figure}[htb]
\begin{center}
\includegraphics[height=4in,width=0.8\textwidth]{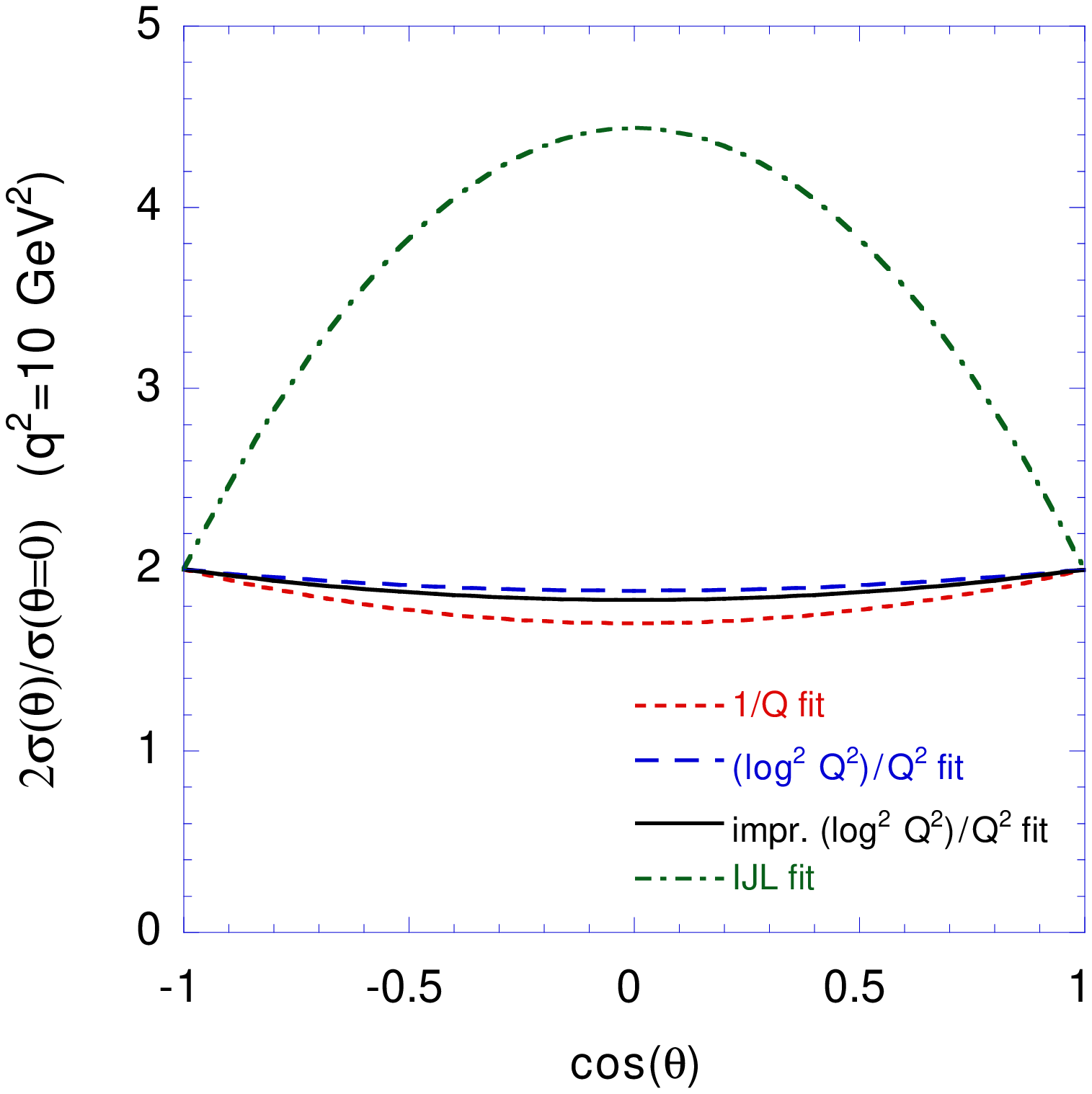}
\end{center}
\caption{The predicted differential cross section
$\sigma(\theta)\equiv d\sigma/d\Omega$. The four curves correspond
to those in Fig.~\ref{figpy}.} \label{fig:diff}
\end{figure}


The magnetic form factor in the IJL model is very small in the 10 to 20
GeV$^2$ region (taking the dipole form for comparison) and has a zero
in the complex plane near $q^2 = 15$ GeV$^2$.   This accounts for much
of the different behavior of the IJL model seen in the polarization
plots.  That the IJL ratio for $G_E/G_M$ is strikingly large even by
the standard set by the other three models also strongly affects the
angular behavior of the differential cross section.  This is witnessed
by Fig.~\ref{fig:diff}, which shows the angular behavior of
$d\sigma/d\Omega$ for $q^2 = 10$ GeV$^2$.  The lower three models are
also showing significant contributions from $G_E$;  at $90^\circ$, the
difference between the curves shown and the value 0.5 is entirely due
to $|G_E|^2$.



\section{Conclusions}                                    \label{theend}


We have discussed how to measure baryon form factors in the timelike
region using polarization observables.  Observing the baryon
polarization in $e^- e^+ \rightarrow B \bar B$ for spin-1/2 baryons $B$
may be the method of choice for determining the magnitude and the phase
of the form factor ratio $G_E/G_M$.  In the spacelike region, one
recalls that at high $Q^2$, the electric form factor makes a small
contribution to the cross section, and the Rosenbluth method of
separating it from the magnetic form factor, by its different angular
dependence, is very sensitive to experimental uncertainties and
radiative corrections~\cite{guichon}.  The more direct method is to use
polarization transfer~\cite{perdrisat, acg81}.  Similarly, in the
timelike case, the angular distribution can be used to isolate
$|G_E|$, but the numerical size of the $G_E$ contribution is small in
many models, whereas two of the three polarization observables are
directly proportional to $G_E$.  Additionally, the phase can only be
measured using polarization.

The normal polarization ${\cal P}_y$ is a single-spin asymmetry and
requires a phase difference between $G_E$ and $G_M$.  It is an example
of how time-reversal-odd observables can be nonzero if final state
interactions give interfering amplitudes different phases. Its analog
in the spacelike case is zero.

A strong current motivation for further baryon form factor study is the
intriguing spacelike JLab data for $F_2/F_1$ or $G_E/G_M$ on the
proton~\cite{perdrisat}.  We have selected a number of fits to this
spacelike data, continued them to the timelike region, and predicted
what size polarizations one may expect to see there.  For the models we
have examined the predicted polarizations are large and distinctive and
should encourage experimental study.


\section*{Acknowledgments}
We wish to thank V.~A.~Karmanov and F.~Iachello for helpful discussions.
This work was supported in part by the Department of Energy
contracts DE-AC03-76SF00515 (S.J.B.), DE-AC05-84ER40150 (S.J.B. and C.E.C),
and DE-FG02-98ER41087 (J.R.H.);
by the National Science Foundation Grant PHY-0245056 (C.E.C);
and by the LG Yonam Foundation (D.S.H.).


\end{document}